\documentclass[%
  reprint,
  superscriptaddress,
  showpacs,
  showkeys,
  amsmath,amssymb,
  pra,
  longbibliography,
  floatfix,
  x11names
]{revtex4-2}

\usepackage{amsmath}
\usepackage{amssymb}
\usepackage{amsthm}
\usepackage{bm} 
\usepackage{graphicx}
\usepackage{tikz}
\usetikzlibrary{decorations.markings,lindenmayersystems}
\usepackage{pgfplots}
\pgfplotsset{compat=1.18}
\usepackage{orcidlink}
\usepackage{hyperref}

\definecolor{cubeface1}{RGB}{180,180,180}  
\definecolor{cubeface2}{RGB}{220,220,220}  
\definecolor{cubeface3}{RGB}{140,140,140}  
\definecolor{cubeedge}{RGB}{80,80,80}      

\newcommand\mengersponge[2][]{
    \pgfkeys{
        /menger/.cd,
        draw edges/.initial=true,
        edge width/.initial=0.1pt,
        .unknown/.style={\pgfkeyscurrentname/.style={#1}}
    }
    \pgfkeys{/menger/.cd, #1}

    \pgfmathsetmacro\level{int(#2 - 1)}
    \ifnum \level = 0 \relax
        \mengercube
    \else
        \begin{scope}[scale=0.333333]
            \foreach \x in {-2,0,2} {
                \foreach \y in {-2,0,2} {
                    \foreach \z in {-2,0,2} {
                        \pgfmathsetmacro{\skipthis}{
                            (\x==0 && \y==0) || (\x==0 && \z==0) || (\y==0 && \z==0) ? 1 : 0
                        }
                        \ifnum\skipthis=0
                            \begin{scope}[shift={(\x,\y,\z)}]
                                \mengersponge[#1]{\level}
                            \end{scope}
                        \fi
                    }
                }
            }
        \end{scope}
    \fi
}

\newcommand\mengercube{
    \filldraw[
        fill=cubeface1,
        draw=cubeedge,
        line width=\pgfkeysvalueof{/menger/edge width}
    ] (1,1,1) -- (1,1,-1) -- (1,-1,-1) -- (1,-1,1) -- cycle;
    \filldraw[
        fill=cubeface2,
        draw=cubeedge,
        line width=\pgfkeysvalueof{/menger/edge width}
    ] (1,1,1) -- (1,1,-1) -- (-1,1,-1) -- (-1,1,1) -- cycle;
    \filldraw[
        fill=cubeface3,
        draw=cubeedge,
        line width=\pgfkeysvalueof{/menger/edge width}
    ] (1,1,1) -- (1,-1,1) -- (-1,-1,1) -- (-1,1,1) -- cycle;
}

\pgfdeclarelindenmayersystem{Koch curve}{
  \rule{F -> F+F--F+F}
}

\pgfdeclarelindenmayersystem{Cantor set}{
  \rule{F -> FfF}
  \rule{f -> fff}
}

\begin{document}

\title{Anti-Gravity from Vacancies in Fractal Space-Time: The Case of a Menger Sponge}

\author{Karl Svozil\,\orcidlink{0000-0001-6554-2802}}
\email{karl.svozil@tuwien.ac.at}
\homepage{http://tph.tuwien.ac.at/~svozil}
\affiliation{Institute for Theoretical Physics, TU Wien, Wiedner Hauptstrasse 8-10/136, 1040 Vienna, Austria}

\date{\today}

\begin{abstract}
We explore the idea that anti-gravity, interpreted as matter-matter repulsion, may emerge as an effective description of spacetime with a reduced local ``substratum density,'' modeled heuristically by vacancies in a fractal lattice. Using the Menger Sponge as a prototypical vacancy-dominated fractal, we motivate a phenomenological identification of a net vacancy parameter with the mass parameter in the Schwarzschild solution. This yields an effective negative-mass Schwarzschild-like metric for embedded observers when vacancies dominate. We analyze curvature, geodesic motion, and energy-condition issues, and we emphasize that our construction is not a microscopic derivation from a specific stress-energy tensor but a re-interpretation of the negative-mass Schwarzschild geometry in terms of fractal vacancies. We discuss conceptual implications, stability challenges, and possible observational signatures.
\end{abstract}


\maketitle

\section{Introduction}
\label{sec:intro}

How would an embedded observer~\cite{toffoli:79,svozil-94} experience~\cite{sv1} motion in a fractal~\cite{falconer1} substratum~\cite{Ord-83}? Throughout, ``embedded'' means an observer who has no kinematic access to any external Euclidean background.

The distinction between intrinsic and extrinsic perspectives is powerfully elucidated in Einstein's ``aether'' metaphor from his 1920 Leiden address~\cite{einstein-aether-en}. He used this analogy to differentiate his concept of a ``gravitational aether'' in general relativity from the discarded notion of a mechanical, light-carrying medium.

Einstein invites us to consider the surface of water upon which waves propagate, and the two vastly different conceptual models or ``mental images''~\cite{hertz-94e} one can construct. First, an extrinsic observer, viewing the system from outside, could track the motion of the underlying substance by tossing small corks onto the water. By following these markers, such an observer gains insight into the movements of the individual particles that constitute the medium.

In contrast, an intrinsic, embedded observer---a ``Flatlander''~\cite{abbott-flatland} confined to the two-dimensional water-air interface---lacks this external vantage point. Bound by their environment, they can only describe the evolving geometric shape of the waves. They perceive the patterns but have no direct access to the substance that supports them.

Einstein's critical insight emerges from a thought experiment: What if there were no ``corks,'' or more precisely, what if---even with the fluid present---no operational means existed to track its individual components? If the only observable phenomenon were the wave's changing form, one would have no grounds to assume the water consists of discrete, movable particles. Even so, Einstein argues, it could still rightfully be called a ``medium.''

This is precisely his reimagined aether: a geometric, gravitational fabric of spacetime that serves as a medium for light and matter, but whose constituent parts---if they exist---cannot be tracked. It represents a ``geometrized'' conception of a medium, devoid of ponderable, mechanical properties from the intrinsic viewpoint.

A modern parallel appears in phenomena like light bulbs or video playback. For an intrinsic observer unable to resolve events faster than the flicker fusion threshold (roughly 50--60 Hz), a discrete sequence of still frames manifests as perfectly continuous motion.

However, the conceptual models and Hertz's ``mental images''~\cite{hertz-94e} we adopt are far from trivial; they fundamentally shape the trajectory of scientific inquiry. To advance our understanding of spacetime, we must interrogate whether its apparent continuity is truly fundamental or merely an emergent property, much like the seamless flow of a video. Progress in physics may stall if we idealize the vacuum as an imponderable continuum, overlooking the possibility that it is, at a deeper level, composed of ``ponderable,'' tangible constituents. Acknowledging this potential could prove essential for unlocking pathways to a more complete theory of spacetime.

We speculate that ``thinning out'' spacetime, in the sense of reducing the density of such hypothetical constituents, leads to an intrinsic perception of negative curvature, manifesting as anti-gravity (repulsion) for observers within the structure. This contrasts with positive curvature from added matter, akin to defects in solids~\cite{Kroner-1958,kroner-1990}. Unlike particle-based interactions, this is envisioned as an effective geometric effect, preserving the equivalence principle at the level of test-body motion.

We build on relativity's geometric ether~\cite{einstein-aether-en,dirac-aether} and induced gravity from quantum fluctuations, as proposed by Sakharov~\cite{Sakharov-67}. Section~\ref{sec:intuitive} provides intuitive analogies, Section~\ref{sec:quant} presents a semi-quantitative, phenomenological construction of a defect-induced Schwarzschild-type metric, Section~\ref{sec:geodesic} analyzes geodesic motion and repulsion, Section~\ref{sec:energy} discusses the associated energy-condition and stability issues, and Section~\ref{sec:discuss} reflects on implications and prospects for testability.

\section{Intuitive Analogies}
\label{sec:intuitive}

\subsection{Fractals and Embedded Observers}

The Menger Sponge~\cite{Menger1926a_reprint,Menger1926b_reprint,Edgar_ClassicsOnFractals},
constructed by iteratively removing the central cube and the six cubes at the center of each face from a unit cube,
possesses a unique combination of properties.
Although its limiting volume is zero, its scaling complexity is captured by a Hausdorff dimension of $\log 20 / \log 3 \approx 2.73$.
In stark contrast, its topological dimension---characterizing its local connectivity---is only~1.
This paradoxical nature is resolved by recognizing that it is not a simple curve,
but a universal curve~\cite{nobeling-1931} (or in Menger's original German, an ``umfassendste eindimensionale Menge'').
As a universal space for the class of compact 1-dimensional metric spaces,
it is complex enough to contain a topologically identical (homeomorphic) subset of any conceivable 1-dimensional object.
This means that any such object---from a simple line segment to a figure-eight, any knot,
or even the Koch curve---can be found as a homeomorphic copy within the Menger Sponge.

As will be argued qualitatively, an effective dimension of less than three implies a ``volume deficit,'' which an embedded observer would perceive as shorter effective geodesics when distances are measured along the remaining substratum. The third iteration of this structure is depicted in Fig.~\ref{fig:menger}. As a universal curve, it is path-connected yet riddled with vacancies within the three-dimensional continuum. This structure raises questions about the nature of continuous motion for an intrinsic observer, echoing Zeno's arrow paradox: apparent continuity might arise not from spatial contiguity alone, but from information about momentum.

Operational metrics on fractals~\cite{Hodel-74,nagata_1985} involve geodesic distances $d_F(p,q)$, the shortest path length in the fractal set $F$~\cite{kigami-2020,gu-2023}. For embedded observers, these are measured by ``counting steps'' in the substratum~\cite{kroner-1985}, projecting fractal geometry onto effective continua~\cite{sv4}.

\begin{figure}[ht]
\centering
\resizebox{.36\textwidth}{!}{
\begin{tikzpicture}[
    x={(0.866cm,-0.5cm)},
    y={(0cm,1cm)},
    z={(-0.866cm,-0.5cm)},
    scale=2
]
    \mengersponge{3}
\end{tikzpicture}
}
\caption{The Menger Sponge, a universal curve with a topological dimension of 1 and a fractal dimension of approximately 2.73,
shown at the third iteration of its construction (gray: material, transparent: vacancy).}
\label{fig:menger}
\end{figure}

\begin{figure}[ht]
\centering
\resizebox{0.36\textwidth}{!}{
  \begin{tikzpicture}
    \def\n{3}
    \draw[l-system={Koch curve, axiom=F, order=\n, angle=60, step=0.5cm}] lindenmayer system;
  \end{tikzpicture}
}
\caption{Extended line which is not a universal curve: Koch Curve at resolution level three.}
\label{fig:Koch}
\end{figure}

\begin{figure}[ht]
\centering
\resizebox{0.4\textwidth}{!}{
  \begin{tikzpicture}
    \def\n{3}
    \draw[
      black,
      line width=2pt,
      l-system={
        Cantor set,
        axiom=F,
        order=\n,
        step=1cm
      }
    ]
    lindenmayer system;
  \end{tikzpicture}
}
\caption{Carved out line: the Cantor set at resolution level three.}
\label{fig:Cantor}
\end{figure}

\subsection{Analogy to Defects in Solids}

Relativity redefined the ether as geometric, not mechanical~\cite{einstein-aether-en}. Quantum fluctuations add ``ponderability'' (e.g., Casimir forces~\cite{Casimir_1948}), and Sakharov viewed gravity as emergent from vacuum elasticity~\cite{Sakharov-67}.

From the onset, the elasticity theory of solids has been linked to the formalism of the general theory of relativity~\cite{schaefer-1953,zaanen-2022}. A geometrical approach to the theory of structural defects in solids by (four-dimensional) continuum mechanics~\cite{Kroner-1958,kroner-1959,Kosevich-1962,Turski-66,kroner-1967,kroner-1975,Kossecka_deWit-77,kroner-1985,kroner-1990,kroner-2001,amari-1968,gunther-1972,Guenther-1979,gunther-1981,gunther-1983,golebiewska-lasota-1979a,golebiewska-lasota-1979b}
effectively~\cite{anderson:73} encodes defects in terms of elastoplasticity.
This formalism is tensor based.

Already Kr\"oner emphasized the
operational, intrinsic viewpoint of embedded observers as being pertinent and fundamental~\cite{kroner-1990}:
``Imagine some crystal being who has just the ability to recognize
crystallographic directions and to count lattice steps along them. Such an
internal observer will not realize deformations from outside, and therefore
will be in a situation analogous to that of the physicist exploring the world.
This physicist clearly has the status of an internal observer.''
Therefore~\cite{kroner-1985}, ``lengths are measured and atoms identified by
counting lattice steps in the three crystallographic directions, then applying Pythagoras' theorem
$
ds^2 = g_{kl}\, dx^k dx^l,
$
where $ds$ is the distance of two atoms with relative position $dx^k$.
$\ldots$
$ds$ $\ldots$ is not the distance obtained by
an external observer by means of a constant scale, but is, rather, the distance found
by an internal observer with the help of the counting procedure''.

The aforementioned analogy between general relativity and the elasticity theory of solids
has led to speculations that dark matter is a solid~\cite{Bucher-PhysRevD.60.043505}.
Kleinert even speculated that a general relativity-type ``crystal gravity'' can be derived for a ``world crystal'' with defects~\cite{kleinert-1987,kleinert-2000,kroner-2001,kleinert-2004}.
In this analogy, the conserved defect tensor can be identified with the Einstein curvature tensor.
The fourth (time) dimension enters because of the dynamics: the movement of defects and the change of the crystal's plastic state~\cite{amari-1968}.

The following metaphor, inspired by Hertz's concept of a ``mental image''~\cite{hertz-94e} and modern condensed matter analogies~\cite{zaanen-2022},
provides an intuitive picture of the concept. It is also relevant to certain quantum gravity models~\cite{Ambjorn2012_review, barkley-2019}:
Suppose you want to envelope a cone with a sheet of paper.
Due to the stiffness of the paper and its nonelasticity this will not be possible.
However, if you allow \emph{adding} stuff (that is, additional paper) to the paper (by cutting and gluing) you can
produce a paper that envelops any kind of surface, including the cone. As a consequence of alignment such an envelope will bow out of the originally flat
paper plane, and produce a non-flat curvature.

This additional stuff will also produce ``longer paths'' from one end of the paper to another,
as crossing over or around the envelope of the cone requires more steps (from ``added stuff'') than on the originally flat surface.
A typical example of such a situation is the Koch curve depicted in Fig.~\ref{fig:Koch}.
Therefore, relative to ``free space'' without defects, adding material (disclination) lengthens geodesics.
Intrinsically, this will be experienced as the tendency of bodies to ``delay separation'' and ``stay together'', thus mimicking attraction.

Conversely, \emph{removing} stuff (vacancies) creates ``shortcuts,'' the capacity to cross points in space
along a ``shorter path''. This will be intrinsically experienced not as contraction but as an expansion,
such that bodies move away from each other faster, suggesting repulsion.
A canonical example of such a cut-out configuration is the Cantor set, depicted in Fig.~\ref{fig:Cantor}.
Likewise, for Menger Sponge observers, vacancies reduce traversable material, accelerating motion intrinsically---perceived as anti-gravity.

\section{Semi-Quantitative Phenomenological Model}
\label{sec:quant}

\subsection{Motivation from the Schwarzschild Metric}

To construct a simple model for a localized defect, we move away from detailed elastic analogies and instead draw on the canonical static, spherically symmetric solution of General Relativity: the Schwarzschild metric. This metric provides the exact description of the spacetime geometry outside a static, spherically symmetric point mass \(M\).

In geometrized units (\(G=c=1\)), the Schwarzschild metric reads
\begin{equation}
g_{\mu\nu} = \text{diag}\left[-\left(1 - \frac{2M}{r}\right), \left(1 - \frac{2M}{r}\right)^{-1}, r^2, r^2 \sin^2 \theta\right].
\label{eq:schwarzschild}
\end{equation}
The single parameter \(M\) has the dimensions of length and encodes the effective gravitational charge of the source.

Our central hypothesis is that a suitably coarse-grained description of a crystalline or fractal defect distribution might be captured by an \emph{effective} mass parameter in such an exterior geometry. Rather than deriving this from a microscopic stress-energy tensor, we adopt a phenomenological ansatz: a net defect strength, determined by vacancies minus interstitials, plays the role of a gravitational charge. The sign of this net charge then determines whether the effective interaction is attractive or repulsive.

\subsection{Defining the Effective Defect Source}
\label{sec:defect-source}

We introduce dimensionless volume fractions
\begin{equation}
N_i,\,N_v \in [0,1],
\end{equation}
representing, respectively, the fractional abundance of interstitials (added ``stuff'') and vacancies (removed ``stuff'') relative to some reference density. We then define the net defect strength
\begin{equation}
S = N_i - N_v.
\end{equation}
This quantity is dimensionless and measures the local balance between added and removed material: \(S>0\) if interstitials dominate, \(S<0\) if vacancies dominate.

To connect this to the fractal structure of the Menger Sponge, we note that at iteration level \(n\) the retained volume fraction is \((20/27)^n\), implying a vacancy fraction
\begin{equation}
v_n = 1 - \left(\frac{20}{27}\right)^n.
\end{equation}
As a first, purely heuristic step, we may identify a vacancy-dominated regime with
\begin{equation}
S \approx -v_n,
\end{equation}
i.e., we assume that the net defect strength \(S\) is proportional to the removed volume fraction in the fractal. This identification is not derived from a detailed microscopic model; it is an ansatz that encodes the intuitive idea that more vacancies correspond to a more negative effective source.

To obtain a quantity with units of length, we introduce a characteristic lattice scale \(L\) (e.g., the side length of the initial cube or a microscopic spacing). We then define an effective source parameter
\begin{equation}
\beta := S L,
\end{equation}
with dimensions of length, analogous to \(2M\) in the Schwarzschild solution. In what follows, we treat \(\beta\) as a phenomenological parameter summarizing the net defect content of the region, and we \emph{assume} that the exterior spacetime can be approximated by a Schwarzschild-type geometry with \(2M\) replaced by \(\beta\).

\subsection{Defect-Induced Metric as Reparameterized Schwarzschild}
\label{2025-menger-gmld}

Implementing this ansatz amounts to the substitution
\begin{equation}
2M \longrightarrow \beta = SL
\end{equation}
in Eq.~\eqref{eq:schwarzschild}. We thus obtain the metric
\begin{equation}
g_{\mu\nu} = \text{diag}\left[-\left(1 - \frac{\beta}{r}\right), \left(1 - \frac{\beta}{r}\right)^{-1}, r^2, r^2 \sin^2 \theta\right].
\label{eq:defect_metric}
\end{equation}
This is \emph{exactly} the Schwarzschild solution with mass parameter
\begin{equation}
2M = \beta = SL, \qquad M = \frac{1}{2} SL.
\end{equation}
No new solution of Einstein's equations is obtained; rather, we re-interpret the mass parameter \(M\) in terms of an effective defect strength \(S\) and a microscopic scale \(L\). The physical content of the metric is therefore identical to that of the standard Schwarzschild metric, but its source parameter is given a different conceptual meaning.

The sign of \(S\) is now crucial:

\begin{itemize}
\item For \(S > 0\) (interstitial-dominated), \(\beta>0\) and the metric coincides with the standard Schwarzschild spacetime of a positive mass \(M>0\).
\item For \(S < 0\) (vacancy-dominated), \(\beta<0\) and the metric corresponds to the exterior geometry of a \emph{negative} mass~\cite{bondi-1957}.
\end{itemize}

In what follows we refer to these two cases as the attractive and repulsive regimes, respectively, emphasizing that this distinction is entirely encoded in the sign of \(\beta = SL\).

\section{Ricci Scalar and Vacuum Character}
\label{sec:ricci}

The Ricci scalar \(R\) is related to the trace of the stress-energy tensor \(T\) via the trace of Einstein's equations (with \(\Lambda=0\)),
\begin{equation}
R = -8\pi T.
\end{equation}
Calculating \(R\) for our metric~\eqref{eq:defect_metric} allows us to characterize the spacetime as vacuum or non-vacuum.

As emphasized in Sec.~\ref{2025-menger-gmld}, the metric~\eqref{eq:defect_metric} is simply the Schwarzschild solution with the identification \(2M=\beta\). The Schwarzschild metric is by construction a vacuum solution of the Einstein field equations for all \(r>0\):
\begin{equation}
R_{\mu\nu} = 0 \quad \text{for} \quad r>0,
\end{equation}
irrespective of the numerical value or sign of \(M\). Consequently,
\begin{equation}
R = g^{\mu\nu} R_{\mu\nu} = 0 \quad \text{for} \quad r>0.
\end{equation}
The exterior spacetime is therefore a vacuum region in which the curvature is entirely encoded in the Weyl tensor (tidal fields), with matter content localized at \(r=0\) in the form of a point mass (positive or negative).

This serves as a consistency check that, given our phenomenological identification \(2M = SL\), the resulting geometry is a legitimate vacuum solution of Einstein's equations outside the source. It does \emph{not} constitute a derivation of this geometry from a specific microscopic defect-based stress-energy tensor. Our model remains phenomenological in that respect.

\section{Geodesic Motion and Repulsion}
\label{sec:geodesic}

To decide whether the effective gravitational field is attractive or repulsive, one must examine the motion of test particles. The simplest and most physical test is to study radial timelike geodesics in the metric~\eqref{eq:defect_metric} and determine the sign of their radial acceleration.

\subsection{Proper Acceleration of Static Observers}

For reference, consider first a static observer at fixed spatial coordinates \((r,\theta,\phi)\). The worldline is
\[
x^\mu(\tau) = (t(\tau), r, \theta, \phi),
\]
with four-velocity \(u^\mu = (u^t,0,0,0)\). Normalization \(g_{\mu\nu}u^\mu u^\nu = -1\) gives
\begin{equation}
(u^t)^2 = \frac{1}{1-\beta/r}.
\end{equation}
The four-acceleration of this static observer is
\begin{equation}
a^\mu = u^\nu \nabla_\nu u^\mu = \Gamma^\mu_{\nu\lambda} u^\nu u^\lambda.
\end{equation}
Only the \(tt\) component contributes to the radial part,
\begin{equation}
a^r = \Gamma^r_{tt} (u^t)^2.
\end{equation}
For the metric~\eqref{eq:defect_metric},
\begin{equation}
g_{tt} = -F,\quad g_{rr} = F^{-1},\quad F(r) = 1 - \frac{\beta}{r},
\end{equation}
one finds
\begin{align}
\Gamma^r_{tt}
&= \frac{1}{2} g^{rr} \left(-\partial_r g_{tt}\right)
= \frac{1}{2} F \, \partial_r F
= \frac{1}{2} F \frac{\beta}{r^2},\\
a^r &= \Gamma^r_{tt}(u^t)^2
= \frac{1}{2} F \frac{\beta}{r^2} \frac{1}{F}
= \frac{\beta}{2r^2}.
\end{align}
Thus the proper radial acceleration required to hold a static position is
\begin{equation}
a^r = \frac{\beta}{2r^2}.
\end{equation}
For \(\beta>0\) this acceleration points outward, meaning a static observer must accelerate upward to resist inward gravitational pull, as in the usual Schwarzschild case. For \(\beta<0\) the sign reverses. However, this describes the acceleration of non-geodesic worldlines maintained in place by some non-gravitational force.

\subsection{Free-Fall Geodesics: Attraction vs.\ Repulsion}

The unambiguous probe of attraction or repulsion is the acceleration of freely falling (geodesic) test particles. For radial timelike geodesics, the line element reduces to
\begin{equation}
ds^2 = -\left(1-\frac{\beta}{r}\right) dt^2
+ \left(1-\frac{\beta}{r}\right)^{-1} dr^2.
\end{equation}
Conservation of energy along the geodesic yields a first integral:
\begin{equation}
E = \left(1-\frac{\beta}{r}\right)\frac{dt}{d\tau},
\label{eq:energy}
\end{equation}
where \(\tau\) is proper time and \(E\) is a constant. Using normalization \(u^\mu u_\mu=-1\) and Eq.~\eqref{eq:energy}, one finds for radial motion
\begin{equation}
\left(\frac{dr}{d\tau}\right)^2 = E^2 - \left(1-\frac{\beta}{r}\right).
\end{equation}
Consider a particle released from rest at radius \(r=r_0\), i.e., with initial condition \((dr/d\tau)|_{\tau=0}=0\). Then
\begin{equation}
E^2 = 1-\frac{\beta}{r_0},
\end{equation}
and near \(\tau=0\) one may expand to find the second derivative of \(r\):
\begin{equation}
\left.\frac{d^2 r}{d\tau^2}\right|_{\tau=0}
= -\frac{\beta}{2 r_0^2}.
\end{equation}
This result is analogous to the Newtonian expression with \(\beta=2M\).

The sign is now decisive:
\begin{itemize}
\item For \(S>0\) (\(\beta>0\)), \(\ddot{r}|_0 = -\beta/(2r_0^2)<0\): the particle accelerates inward; the field is attractive.
\item For \(S<0\) (\(\beta<0\)), \(\ddot{r}|_0 = -\beta/(2r_0^2)>0\): the particle accelerates outward; the field is repulsive.
\end{itemize}
Thus, when vacancies dominate (\(N_v > N_i\), hence \(S<0\)), the effective metric~\eqref{eq:defect_metric} describes matter-matter repulsion in the sense of free-fall geodesics. This matches the qualitative picture of vacancies creating intrinsic ``shortcuts'', and it is fully equivalent, at the level of the exterior geometry, to the well-known negative-mass Schwarzschild solution~\cite{bondi-1957}.

\section{Energy Conditions and Exotic Matter}
\label{sec:energy}

Energy conditions are criteria imposed on the stress-energy tensor \(T_{\mu\nu}\) to capture the idea of physically reasonable matter. They constrain combinations such as \(T_{\mu\nu}u^\mu u^\nu\) (energy density) and its variants for null vectors. We briefly review how these conditions apply to our defect-reinterpreted Schwarzschild metric.

\subsection{Vacuum Exterior ($r > 0$)}

As established in Sec.~\ref{sec:ricci}, the spacetime described by Eq.~\eqref{eq:defect_metric} is a vacuum solution for all \(r>0\), in the strict GR sense:
\begin{equation}
T_{\mu\nu} = 0 \quad (r>0).
\end{equation}
All local energy conditions are then trivially satisfied in the vacuum region, e.g.,
\begin{equation}
T_{\mu\nu} k^\mu k^\nu = 0 \ge 0
\end{equation}
for any null vector \(k^\mu\) (Null Energy Condition), and similarly for timelike vectors. The vacuum itself is not exotic; the exoticity, if any, must be localized at the source.

\subsection{Strong Energy Condition and Repulsion}

The Strong Energy Condition (SEC) can be recast geometrically through Einstein's equations as
\begin{equation}
R_{\mu\nu}u^\mu u^\nu \ge 0
\end{equation}
for any timelike vector \(u^\mu\). Combined with the Raychaudhuri equation, this condition ensures that gravity is, on average, attractive: timelike geodesic congruences tend to focus.

In our case, \(R_{\mu\nu}=0\) for \(r>0\), so
\begin{equation}
R_{\mu\nu}u^\mu u^\nu = 0.
\end{equation}
The vacuum region therefore \emph{saturates} the SEC: it neither contributes to focusing nor to defocusing via the Ricci tensor. All gravitational effects in the vacuum arise from the Weyl tensor (tidal fields), not from local matter content.

The global character of the field (attractive vs.\ repulsive) is encoded in the sign of the mass parameter \(M=\beta/2\). For \(M>0\) we have the usual inward attraction, while for \(M<0\) we have outward repulsion as shown in Sec.~\ref{sec:geodesic}. This difference originates in the effective source at \(r=0\).

\subsection{Nature of the Source at the Origin}

The exterior geometry of Eq.~\eqref{eq:defect_metric} can be thought of as produced by a point-like source at \(r=0\) with total gravitational mass
\begin{equation}
M = \frac{1}{2} SL.
\end{equation}
The sign of \(S\) determines whether this mass is positive or negative.

\begin{itemize}
\item \textbf{Attractive regime (\(S>0\)):} The source has positive mass and can in principle be modeled by conventional matter distributions satisfying the usual energy conditions in an appropriate limit.
\item \textbf{Repulsive regime (\(S<0\)):} The source must have negative mass and therefore correspond to exotic matter, violating the averaged forms of the energy conditions. As emphasized long ago by Bondi~\cite{bondi-1957}, such negative-mass sources are associated with run-away instabilities (e.g., in a positive/negative mass pair) and other conceptual issues.
\end{itemize}

Our identification of \(M\) with a defect parameter \(S\) does not eliminate these fundamental problems; it merely reinterprets the negative mass in terms of an effective vacancy-dominated region. From the standpoint of classical GR, the repulsive model still requires an exotic source.

\section{Discussion}
\label{sec:discuss}

We have presented a phenomenological framework in which anti-gravity, understood as matter-matter repulsion in the sense of free-fall geodesics, is associated with vacancy-dominated regions of an underlying substratum, exemplified by the Menger Sponge fractal. The key steps are:

\begin{enumerate}
\item Introduce a net defect strength \(S = N_i - N_v\), with \(N_i\) and \(N_v\) interpreted as volume fractions of interstitials and vacancies, respectively.
\item Identify a characteristic microscopic scale \(L\), and define a phenomenological effective source parameter \(\beta = SL\) with dimensions of length.
\item Assume that on large scales the exterior geometry of a localized defect region is well approximated by a Schwarzschild-type metric with \(2M = \beta\), i.e., the line element~\eqref{eq:defect_metric}.
\item Analyze geodesic motion and show that for \(S<0\) the resulting dynamics corresponds to that of a negative-mass Schwarzschild spacetime, exhibiting repulsive free-fall acceleration.
\end{enumerate}

Mathematically, our construction does not yield a new solution of the Einstein field equations; it is a reparameterization of the Schwarzschild solution with a mass parameter \(M=\tfrac{1}{2}SL\). The innovation lies in the proposed interpretation of \(M\) in terms of vacancy-dominated fractal structures. The solid-defect analogy---where added material lengthens geodesics (attraction) and removed material shortens them (repulsion)---provides an intuitive bridge between condensed matter physics and general relativity.

This framework fits conceptually with Sakharov's idea of induced gravity~\cite{Sakharov-67} and with attempts to describe defects in a crystalline vacuum~\cite{kleinert-1987,kleinert-2000,kroner-2001,kleinert-2004}. However, several important caveats must be emphasized:

\begin{itemize}
\item The identification \(2M = SL\) is an \emph{ansatz}, not a derivation from a microscopic stress-energy tensor built from a Menger-sponge or other fractal distribution. We have not solved Einstein's equations with a true fractal source.
\item The model at the level of the exterior geometry is \emph{equivalent} to the standard negative-mass Schwarzschild solution~\cite{bondi-1957}. The fractal-vacancy picture re-labels the mass parameter but does not, as yet, provide a mechanism that resolves the well-known stability and energy-condition problems of negative mass.
\item The empirical status of fractal spacetime remains speculative. While fractals appear in many effective descriptions and in some approaches to quantum gravity~\cite{Ambjorn2012_review,barkley-2019}, there is currently no direct observational evidence that spacetime itself has a Menger-sponge-like vacancy structure on any scale.
\end{itemize}

Despite these limitations, the model suggests several directions for further work:

\begin{enumerate}
\item \textbf{From toy fractal to effective density profile:} One could construct spherically symmetric effective density profiles that mimic the radial scaling of a fractal (e.g., \(\rho(r)\propto r^{D-3}\) for an effective fractal dimension \(D<3\)) and solve the Einstein equations for such matter distributions. This would provide a more direct link between fractal geometry and spacetime curvature.
\item \textbf{Observational signatures:} Negative-mass Schwarzschild spacetimes predict defocusing gravitational lensing of light, modified Shapiro time delays, and altered orbital dynamics. Even in the absence of a concrete microscopic model, one can compute such effects explicitly and compare with lensing surveys or timing observations to place bounds on the existence of compact repulsive objects.
\item \textbf{Quantum and stability issues:} The run-away behavior and energy-condition violations associated with negative mass are serious problems. Embedding the vacancy interpretation in a quantum-field-theoretic or quantum-gravity context could, in principle, lead to stabilized effective negative-energy configurations (e.g., Casimir-like effects~\cite{bekenstein-2013,costa-2022,kontou-2020}). At present, however, this remains highly speculative~\cite{Santiago22-PhysRevD.105.064038}.
\end{enumerate}

In summary, we have shown that if one accepts the phenomenological identification \(2M = SL\), then vacancy-dominated fractal regions can be reinterpreted as effective negative-mass sources producing repulsive gravitational fields in the precise sense of geodesic motion. This does not yet constitute a complete physical theory of anti-gravity, but it offers a concrete, geometrically motivated way to think about repulsion as arising from ``less-than-space'' rather than from additional exotic particles. Future progress will require deriving such effective parameters from an underlying microphysics, resolving the associated energy-condition violations and instabilities, and confronting the resulting models with observations.

\begin{acknowledgments}
This text was partially created and revised with assistance from one or more of the following large language models: Grok4-0709, Gemini 2.5 Pro, o3-2025-04-16, claude-sonnet-4-20250514-thinking-32k, GLM-4.5. All content, ideas, and prompts were provided by the author.
This research was funded in whole or in part by the Austrian Science Fund (FWF) Grant DOI: 10.55776/PIN5424624.
The author acknowledges TU Wien Bibliothek for financial support through its Open Access Funding Programme.
\end{acknowledgments}

\bibliography{svozil}

\end{document}